# THE GAMMA-RAY BURSTS AND CORE-COLLAPSE SUPERNOVAE - GLOBAL STAR FORMING RATE PEAKS AT LARGE REDSHIFTS

V. V. SOKOLOV

*Special Astrophysical Observatory of the Russian Academy of Sciences (SAO RAS), Nizhnij Arkhyz 369167, Russia*

This is a brief review on the first Gamma-Ray Bursts (GRB) optical identifications – GRB host galaxies and Star Forming Rate (SFR) at relatively small redshifts ($z$), on the metallicities of GRB hosts and the similarities and differences between GRB hosts and galaxies at larger $z$, and on the SFR and GRB rate (GRBR) at the high $z$. Evidences of a direct connection between long-duration GRBs and massive stars explosions (like Core-Collapse Super-Novae – CCSNe) are presented. Is there a fast decrease in SFR up to z ~10? Some unsolved problems related to GRBs are discussed: about the high-$z$ GRB host galaxies, the high-$z$ CCSN-GRB connection, and possible new crucial cosmological tests at high $z$.

## 1   Introduction

Gamma-ray bursts (GRBs) are the brief (~0.01-100$s$), intense flashes of γ-rays (mostly sub-MeV) with enormous electromagnetic energy release up to ~$10^{51}$-$10^{54}$ ergs. The rapid temporal variability, δT <10 msec, observed in GRBs implies *compact* sources with a size smaller than cδT < 3000 km. After the first optical identification of GRBs in 1997 [1, 2] they became right away a new direction in the study of the universe at large redshifts. In particular, for GRB 090423 the redshift $z$ = 8.2 was spectroscopically confirmed [3, 4], and for GRB 090429B Cucchiara et al. [5] have estimated the photometric redshift $z \sim$ 9.4. So, at the present time the general state of the GRB problem and progress in this field could be categorized in the following way: 1) GRBs belong to the most distant objects with measurable redshifts in the universe, 2) GRBs are related with star formation in distant (and very distant, $z \sim$ 10) galaxies, 3) GRBs and their afterglows allow us seeing the most distant massive star explosions at the end of their evolution, 4) This is confirmed by observations for long-duration GRBs, but, most probably, short GRBs (<1$s$) are also related to some very old compact objects formed by evolution of identical massive stars.

## 1   Optical identification: the first GRB host galaxies and massive SFR

X-ray and optical afterglows were observed for the very first time [1, 2] for GRB 970228 by the Italian-Dutch satellite BeppoSAX [6] thanks to the fast and accurate positioning of GRB obtained through combined capabilities of the GRB Monitor and Wide Field Cameras onboard this famous satellite. The next GRB afterglow (GRB 970508) was observed optically already with BTA (the 6-m telescope of SAO RAS) in standard BVRcIc bands in October-December 1997 up to January 1998. The first results of photometry of the afterglow and three nearby galaxies were presented in [7]. Further multi-wavelength observations of GRBs have confirmed that a significant fraction of long-duration GRBs is associated with collapse of short-lived massive (~ $30M_\odot$) stars [8, 9]. The regions of massive star-forming are seen well in the UV part of the galaxy spectra – this is the light of massive stars in the GRB hosts. The comparison of photometrical properties of star-forming galaxies and the first sample of the GRB host galaxies are presented in [10, 11, 12].

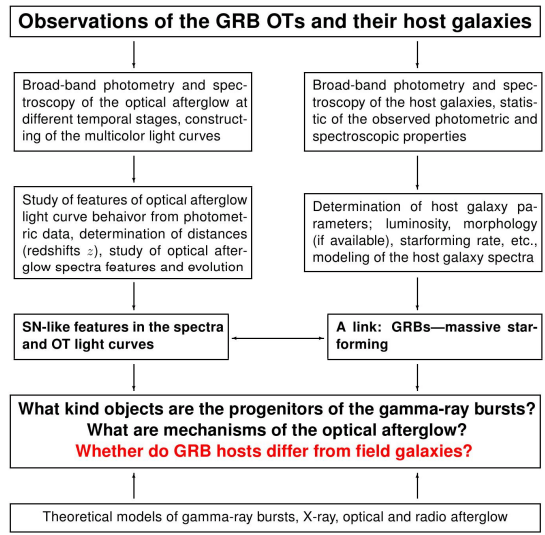

Figure 1. The astronomy of GRBs with BTA since 1998: OT is an optical transient source connected with a GRB.

The fast localization of GRBs by BeppoSAX and their ground-based follow-up optical observations (+ redshift measurements) have established the relation between GRBs and these distant galaxies (the GRB hosts) located in sites of the faded transients (OT). This was essentially *the first stage of optical identification* with star-forming galaxies – the objects with more or less distinct properties in contrast to properties of emission from the GRB optical afterglow itself. That is the study of these objects (the broad-band photometry and spectroscopy of GRB hosts, statistics of observed photometric and spectroscopic properties, and so on) that launched the GRB astronomy and with BTA as long ago as in 1998 (see Fig. 1). Researchers studying GRB optical counterparts were interested at that time mostly in the question: do GRB hosts differ from field galaxies or not? The study of physical properties of the GRB hosts permits determining differences among others galaxies (*in the same fields*, as it was done for GRB 970508 ($z$ = 0.8349) [10, 11]) with massive star-forming ones, which gives us a key to the understanding of conditions in which GRB progenitor objects are born, evolve and end their life.

But the most distant host galaxies can be often observed only photometrically. In these cases, such physical properties as SFR, intrinsic extinction, ages, masses and metallicities (see Table 1) can be estimated by the population synthesis modeling of energy distributions in galaxy spectra. For example, in the case of the distant host of GRB 980703 ($z$ = 0.9662) the observed deficit in the B-band could be explained by dust extinction excess near 2200Å [11], which is the characteristic *dust extinction* for our Galaxy. This extinction was confirmed later [13] for even *more distant* GRB hosts with larger $z$: e.g. the GRB 070802 ($z$ = 2.4541) which was observed with VLT/FORS2, and GRB 080607 ($z$ = 3.0368) with the Keck spectrum. The 2175Å dust extinction feature is clearly seen in the afterglow spectra (see e.g. Fig. A.2 in [13]).

For GRB hosts with the very different *absolute* magnitudes (GRB 970508, $M$ = -18.62 and GRB 980703, $M$ = -21.27) we made the continuum Spectral Energy Distributions (SED) modeling, making use of the spectra and our photometry. The broad-band photometric SEDs describe well the spectra of the respective host galaxies [12]. The identical method of the broad-band SEDs fitting (determination of $z$, luminosity, stellar mass, age, metallicity and SFR as it is in Table 1) is now widely used in studies of very distant and faint (with observed magnitudes > 28) galaxies involving results of the broad-band photometry in infra-red frequencies. (Good examples of the best-fit stellar population models for $z$ = 7-8 galaxies are shown in [14].)

Table 1. The main parameters for two GRB hosts from the stellar population modeling.

| Host | $z$ | Observ. R mag. | Absolute mag. ($M_{Brest}$) | Stellar mass ($M_\odot$) | Age | Metallicity | SFR |
|---|---|---|---|---|---|---|---|
| GRB 970508 | 0.8349 | 25.0 | -18.62 | $3.48 \cdot 10^8$ | 160 Myr | $0.1 Z_\odot$ | $14 M_\odot yr^{-1}$ |
| GRB 980703 | 0.9662 | 22.3 | -21.27 | $3.72 \cdot 10^{10}$ | 6 Gyr | $Z_\odot$ | $20 M_\odot yr^{-1}$ |

To summarize results of the GRB host modeling [11,12], we can conclude the following: (i) The broad-band flux spectra of GRB hosts are well fitted by SEDs of *local star-burst galaxies*. (ii) The UV part of GRB host galaxy SEDs are properly described by models with *young burst* star formation, and for $z \sim 1$, in the optical wavebands we observe only star-forming regions in GRB galaxies, because the *massive stars* dominate the rest-frame UV part of spectrum. (iii) GRB host galaxies seem to be identical to usual galaxies at identical $z$ (see Fig. 2).

So, we have concluded that long-duration GRBs seem to be closely related to vigorous (massive) star formation in their host galaxies. It should be noted that the SFR in host galaxies is unlikely to be much higher than in galaxies at identical $z$. At this $z$ the mean star formation rate is $\sim 20 - 60 M_\odot yr^{-1}$ (see also [15]). For these reasons we conclude that GRB host galaxies seem to be similar to field galaxies (at identical $z$ [11, 12]). At present, we have an *independent* confirmation by Savaglio et al. [16]: GRBs are identified within ordinary galaxies indeed. The paper was based on the GRB Host Studies public archive collecting observed quantities of 32 GRB host galaxies, i.e. about half of the total number of GRBs with redshifts known by January 2006. The authors present some preliminary statistical analysis of the sample, e.g. the total stellar mass, metallicity and the SFR for the hosts. The total stellar mass and the metallicity for a subsample of 7 GRB hosts at $0.4 < z < 1$ are consistent with the mass-metallicity relation found for normal star-forming galaxies in the identical redshift interval.

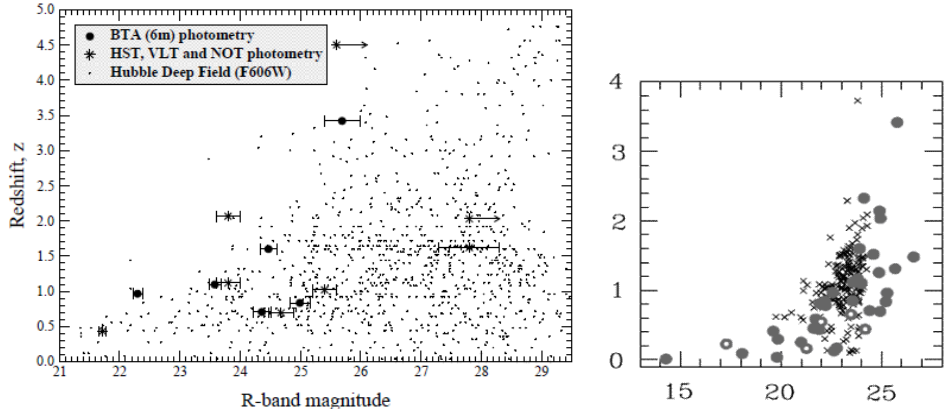

Figure 2. Left: The observed *R*-band magnitudes versus spectroscopic *z* for the first 12 GRB host galaxies [11,12]. The BTA magnitudes are marked with circles, while asterisks refer to the results of other authors. The catalog of the HDF F606W magnitudes and *photometric* redshifts were used (points). Right: The observed magnitudes of a larger sample of GRB hosts *vs.* spectroscopic *z* from the paper by Savaglio et al. [17]: GRB hosts (filled circles) and spectroscopic *z* from Gemini Deep Survey field galaxies (crosses). The empty circles are short-GRB hosts.

The observed magnitudes of more than 30 GRB host galaxies as a function of *z* (Fig. 2, right) were shown in the paper by Savaglio et al. [17] for the GRB hosts and Gemini Deep Survey field galaxies. In this later study (see in [17] and references therein) the authors formulated more definite conclusions but for a sample of GRB hosts with *larger z*: There is no clear indication that GRB host galaxies belong to a special population. Their properties are similar to those expected for normal star-forming galaxies, from the local to the most distant universe. Combining these with the results for GRB hosts with $z \sim 1$, we see no significant evolution of metallicity in GRB hosts in the interval $0 < z < 6$. This fact implies that GRB hosts do not differ substantially from the typical galaxy population. The low, sub-solar metallicity were found in many studies [17, 18, 19] (and references therein). But it does not necessarily mean that GRBs occur in special low-metallicity galaxies only. The low metallicities in GRB hosts (like in the Magellanic Clouds) are consequences of the high massive star formation. So, the same SFR *appears to be the primary parameter to generate GRB events*.

## 2 The direct connection between long-duration GRBs and massive stars (GRB–CCSN)

Thus, from the aforesaid it follows that there are multiple long lines of evidence that long-duration (~ 2*s* - 100*s*) GRBs are associated with collapse of massive stars, occurring in regions of active star formation embedded in dense clouds of dust and gas. It has been established also that there are *direct* connection between GRBs and massive progenitor stars of CCSNe. Identification of GRB 980425 with CCSN SN1998bw [20] is considered to be the first one. Correspondingly, since 1998 another direction (parallel to the study of GRB hosts, see Fig. 1) is the observation of mysterious optical transients (OTs) connected with GRBs [21, 22]. From the very beginning the main aim of these observations was the study of photometric features in GRB optical afterglow light curves connected with underlying SNe [23].

Some of the nearby ($z < 0.7$) GRBs have shown re-brightening and flattening in their late optical afterglows, which have been interpreted as emergence of the underlying SN [24]. But usually the data for larger $z$ are not of sufficient quality. The SN is too faint to search for such features in the late-time afterglow light curves. In addition, the re-brightening in late optical afterglows for the large $z$ is to be observable already in near-IR. This extra light could be modeled well by an SN component, peaking at $(1+z)\times(15-20)$ days after a burst. This, together with the spectral confirmation of SN light in the afterglows of GRB 021211, GRB 030329, and GRB 031203, further supports the view that, in fact, many long-duration GRBs show SN bumps in their late-time optical afterglows. The fact that a strong late-time bump was also found for (X-ray flash) XRF 030723 [25] and for XRF 020903 [26] might indicate that this conclusion is true for XRFs as well. So, a systematic study of GRB afterglows suggests that *all* long-duration GRBs are associated with SNe [24]. By now many GRBs with spectroscopic confirmed SNe are known: see also the review in [27]. Figure 3 shows some spectroscopic observational results of GRB/XRF 060218/SN 2006aj ($z = 0.0335$) and XRF 080109/SN 2008D ($z = 0.0065$) from GRB/XRF list observed with BTA.

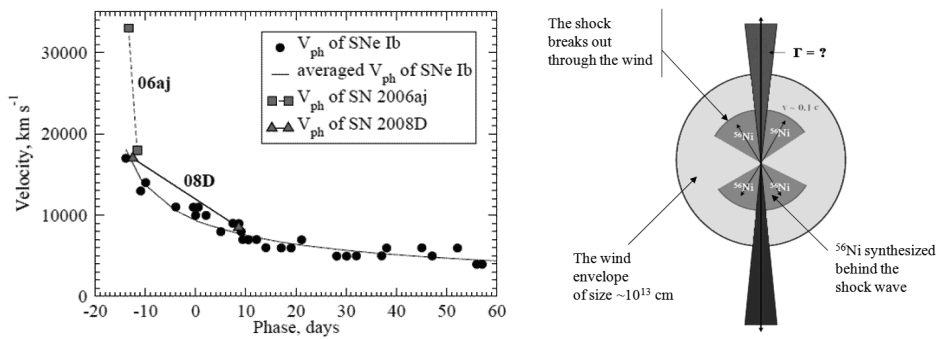

Figure 3. Left – Velocities at SNe photospheres versus time after GRB/XRF. The line is a power-law fit to the data as inferred from Fe II lines (filled circles) for the core collapse SNe Ib type (see in [28]). Squares (GRB/XRF 060218/SN 2006aj) and triangles (XRF 080109/SN 2008D) are photospheric velocities, inferred from the BTA spectra [29, 30]. Right – The schematic model of GRB/SN progenitor asymmetric explosion [31].

Thereby, a long duration GRB can be the beginning of a CCSN, and GRB itself is a signal allowing us catching these massive core-collapse SN at the *very beginning* of the explosion. As a matter of fact, observing GRB, we observe relativistic collapse of a massive star core at the end of its evolution.

At least, it seems that a closer GRB can reveal more features of SN. So, nearby GRB afterglows can show more spectroscopic signs of SNe. Though the phenomenon (GRB per se) is unusual, but the source object (SN) is known for a long time already. (This is similar to the situation with GRB host galaxies, as was seen before.) A popular concept of the relation between long-duration GRBs and CCSNe is shown in Fig. 3 (the picture from [31]). The γ-ray emission in a narrow cone (from a central GRB source) is observed along the SN explosion axis, but closer to the equatorial plane we can observe mainly only an almost isotropic X-ray flash (XRF) related to a shock break-out effect [31]. This can explain why most local SNe do not show any GRB, though they can show a powerful and short duration XRF phase (as was in the case of XRF 080109/SN2008D with $z = 0.0065$). The probability of getting into a narrow beam of γ-ray quanta decreases with the Lorentz factor Γ. (! But it rapidly rises with the increase of volumes in which GRBs are observed, i.e. with the increase of $z$.)

In the context of aforesaid about GRB host galaxies, the GRB-SN connection could be considered as *the second result of identification* of GRBs. The question "What kind of objects are the progenitors of GRBs?" (Fig. 1) becomes especially important at very high redshifts $z \sim 10$ or more.

## 3  The GRB rate and SFR at $z$~10 – on the global SFR peaks at high redshifts

Thus, the long-duration GRBs with the peak emission at sub-MeV energies (*where extinction is not an issue*) are explosions associated with the core collapse of short-lived massive stars (~ $30M_\odot$). On the other hand, the massive stars death rate (CCSNe rate) must resemble their formation rate – the massive stars SFR. If the SFR is directly proportional to the *GRB formation rate* (SFR ~ GRBR) then the GRB rate (GRBR) could be used as a potential tracer of the massive SFR in the distant universe [32].

And if the correlation SFR ~ GRBR exists up to high $z$, the questions arise: Is the faster decrease of SRF at $z > 4$ observed indeed, which is to be observed in cosmological models? Is there any difference between GRBR and SFR beyond $z \sim 4$? Some comments on SFR in galaxies with large $z$ (which serve as powerful probes of SFR studies at highest $z$) are given below.

As long-duration GRBs are associated with massive stars, for the reasons of star formation, they (GRBs) are good candidates to study the SFR density itself. GRB 090423 at $z = 8.2$ (see also [5] on the redshift $z = 9.4$ for GRB 090429B) has further extended the $z$ interval where the estimates of SFR evolution could be done; this distance regime was never explored before. Kistler et al. [33] have compared SFR for different field galaxy samples with those derived using GRBs. It is based on the idea [32] that the GRBR in galaxies is proportional to SFR and that the ratio GRBR/SFR does not change with $z$. The normalization of GRB SFR density is done by taking the SFR density value at low $z$ for which the densities of the GRBR are measured most precisely [32].

The principal thing is that SFR inferred from GRBs *could remain high*, at least up to $z \sim 9$ [33 and references therein]. The agreement with direct observations, corrected for galaxies below detection thresholds, suggests that the GRB-based estimates incorporate the bulk of high-$z$ star formation down to the faint or low luminosity galaxies (at least down to $M \approx$ -10 mag.). At $z \sim 8$, GRB SFR density is consistent (more or less) with others measurements after accounting for *unseen* galaxies. This implies that not all star-forming galaxies for given $z$ are currently taken into account in deep surveys. GRBs provide the strong contribution to the SFR from large number of small/faint/unseen galaxies. So, the typical GRB hosts at high $z$ might be exactly these small star forming galaxies. Finally, that is what explains why *no steep drop or fast decrease exists in the SFR density inferred from GRBs up to $z \sim 9$* [33].

As a matter of fact, the identical conclusion was obtained in the study by Robertson & Ellis [34]. Here the authors tried to take into account a possible dependence of the GRBR/SFR relation on $z$. Implications of GRB-derived estimates for the high-$z$ SFR density are shown in Fig. 4a. If the GRB rate to SFR ratio (GRBR/SFR) evolves weakly beyond $z > 4$ (see for the details in [34]), the rate of discovery of high-$z$ GRBs already implies a SFR density much larger than that inferred from UV-selected galaxies, as was seen also from [33]. A faster evolution (GRBR $\sim$ SFR$\times(1+z)^{1.5}$, see Fig. 4a) would be needed to force agreement. Parameterized star formation histories consistent with the GRB-derived star formation histories in the *constant and low metallicity* star formation models are shown as *black thick lines*.

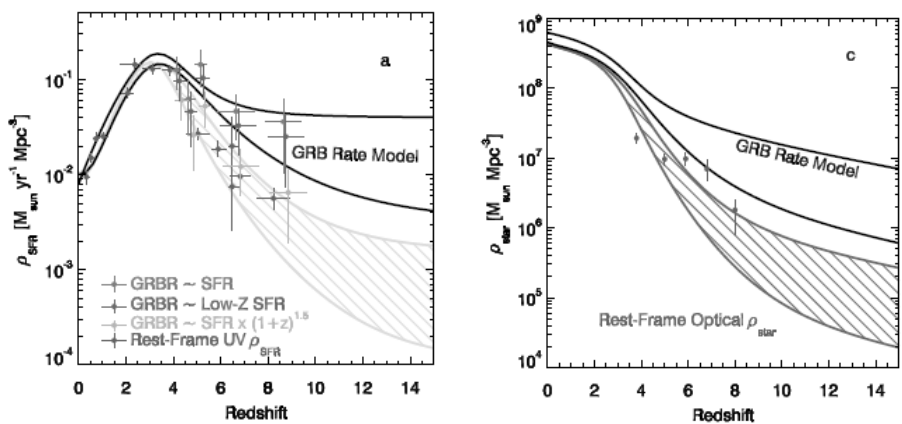

Figure 4. The cosmic SFR density (a) and stellar mass density (c) taken from [34] – see text.

The stellar density (Fig. 4c) is simply determined by *the integral* of the previous SFR density (Fig. 4a). The stellar mass density to $z \sim 9$ is shown as gray points with error bars [35], with the associated models [34] (gray hatched region). The *black thick lines* in Fig. 4c show the stellar mass density implied by parameterizations of the GRB-derived star formation rate, which clearly exceed the models stellar mass density at all $z$. Such SFR density implied by the high-$z$ GRB rate appears unphysical in that it *overproduces the observed stellar mass density at $z \geq 5$*. In any case, these results

48

already affect the choice of cosmological parameters and star formation models of the cosmic star formation history in ΛCDM cosmological simulations (see more in [38]).

**4  Conclusions**

1. Investigation of GRB hosts is the first result of the GRB optical identification in 2001 with known objects: GRBs are identified with ordinary (or the most numerous in the universe at any z) galaxies up to ~ 28 mag. or fainter. The GRB hosts are not special, but normal, frequently faint, star-forming galaxies most abundant, and they are detected at any z just because a GRB event has occurred. So, GRB hosts do not differ from the star-forming galaxies at small $z$ – neither in colors, nor in spectra, SFRs, and metallicities. These are generally star-forming galaxies ("ordinary" for their $z$) constituting the base of deep surveys [9, 36, 37, 11, 16, 17].
2. Now long-duration GRBs could be identified with ordinary massive CCSNe. So, we have the massive star-formation in GRB hosts and massive star explosions in form of CCSNe/GRB. The search for differences between nearby SNe identified with GRBs and distant SNe which are to be identified with GRBs could be an additional observational cosmological test for $z \sim 10$. We can ask a question analogous to that of GRB hosts identified in 2001: Do GRB-SNe differ from usual SNe? Generally, at what redshifts the SNe show different properties from the local CCSNe population?
3. As the universe is transparent to γ-rays up to $z \sim 10\text{-}20$ and more, a new branch of observational cosmology has come-up. The GRBs and their hosts themselves are considered as tools to study processes of star-formation at cosmological distances up to $z \sim 10$ and more. Irrespective of specific models of the GRB phenomenon, it can be said that while observing GRBs, we observe SNe, which are always related to the relativistic collapse of massive stellar cores in very distant galaxies. Up to what redshift ($z > 10\text{-}50?$) are GRBs and massive CCSNe observable? This question would serve as a driver for the future cosmological tests using GRBs-CCSNe.

**References**

1. E. Costa, F. Frontera, J. Heise, et al., *Nature* **387**, 783 (1997).
2. J. van Paradijs, P. J. Groot, T. Galama, et al., *Nature*, **386**, 686 (1997).
3. N. Tanvir, D. B. Fox, A. J. Levan, et al., *Nature*, **461**, 1254 (2009).
4. R. Salvaterra, M. Della Valle, S. Campana, et al., *Nature*, **461**, 1258 (2009).
5. A. Cucchiara, A. J. Levan, D. B. Fox, et al., *ApJ* **736**, 7 (2011).
6. G. Boella, R. C. Butler, G. C. Perola, et al., *A&ASS* **122**, 299 (1997).
7. S. V. Zharikov, V. V. Sokolov, and Yu. V. Baryshev, *A&A*, **337**, 356, arXiv:9804309 (1998).
8. D. W. Hogg, A. S. Fruchter, *ApJ* **520**, 54 (1999).
9. J. S. Bloom, S. G. Djorgovski, S. R. Kulkarni, *ApJ* **554**, 678 (2001).
10. V. V. Sokolov, S. V. Zharikov, Y. V. Baryshev, et al., *A&A,* **344**, 43, arXiv:9809111 (1999).
11. V. V. Sokolov, T. Fatkhullin, A. J. Castro-Tirado, et al., *A&A*, **372**, 438, arXiv:0104102 (2001).

12. V. V. Sokolov, T. Fatkhullin, *Bull. Spec. Astrophys. Obs.*, **51**, 48 (2001).
13. T. Zafar, D. Watson, J. P. U. Fynbo, et al., 2011, *A&A*, 532, 143
14. I. Labbé, V. González, R. J. Bouwens, et al., *ApJL*, arXiv:0911.1356v5 (2009).
15. A. W. Blain, P. Natarajan, *MNRAS* **312**, L39 (2000).
16. S. Savaglio, K. Glazebrook., D. Le Borgne, Gamma-ray bursts in Swift era, 16th Maryland Astrophysics Conference, AIP Vol. 836, p. 540, arXiv:0601528 (2006).
17. S. Savaglio, K. Glazebrook, D. Le Borgne, *ApJ*, **691**, 182 (2009).
18. E M. Levesque, L. J. Kewley, E. Berger, et al., *AJ*, **140**, 1557 (2010).
19. Jing-Meng Hao, Ye-Fei Yuan, ApJ, 772, 42, arXiv:1305.5165 (2013)
20. T. J. Galama, P. J. Groot, J. van Paradijs, et al., *ApJ* **497**, L13 (1998).
21. V. V. Sokolov, A. I. Kopylov, S. V. Zharikov, et al., *A&A*, **334**, 117, arXiv:9802341 (1998).
22. S. V. Zharikov, V. V. Sokolov, *A&ASS*, 138, 485, arXiv:9904160 (1999).
23. V. V. Sokolov, in Proc. "Gamma-Ray Bursts in the Afterglow Era: 2nd Workshop", eds. Costa E. et al., ESO Astrophysics Symposia, Berlin: Springer Verlag, p. 136, arXiv:0102492 (2001).
24. A. Zeh, S. Klose, D. H. Hartmann, in Proc of the 22nd Texas Symposium on Relativistic Astrophysics at Stanford. Stanford California, Dec. 13-17, 2004. ed. Chen P, et al., arXiv:0503311 (2004).
25. J. U. P. Fynbo, J. Sollerman, J. Hjorth, et al., *ApJ* **609**, 962 (2004).
26. A. M. Soderberg, S. R. Kulkarni, D. B. Fox, et al., *ApJ*, **627**, 877 (2005).
27. N. Gehrels & S. Razzaque, , arXive:1301.0840 (2013)
28. D. Branch, S. Benetti, D. Kasen, et al., *ApJ* **566**, 1005 (2002).
29. E. Sonbas, A. S. Moskvitin, T. A. Fatkhullin, et al., *Astrophys. Bull.,* **63**, 228, arXiv:0805.2657 (2008)
30. A. S. Moskvitin, E. Sonbas, V. V. Sokolov, et al., *Astrophys. Bull.,* **65**, 132, arXiv:1004.2633 (2010).
31. S. Woosley, A. Heger, *AIP Conf.Proc.*, **836**, 398 (2006).
32. E. Ramirez-Ruiz, E. E. Fenimore, N. Trentham, arXiv: astro-ph/0010588, talk given at the CAPP2000 Conference on Cosmology and Particle Physics, Verbier, Switzerland, eds. J. Garcia-Bellido, R. Durrer and M. Shaposhnikov, (AIP, 2001).
33. M. D. Kistler, H. Yksel, and A.M.Hopkins, arXiv:1305.1630.
34. B. E. Robertson & R. S. Ellis, Astrophys. J. 744, 95 (2012)
35. V. González et al. 2011, ApJ, 735, L34
36. S. G. Djorgovski, S. R. Kulkarni, J. S. Bloom, et al., invited review in proc. "Gamma-Ray Bursts in the Afterglow Era: 2nd Workshop", eds. Costa E. et al., ESO Astrophysics Symposia, Berlin: Springer Verlag, p. 218 (arXiv:0107535) (2001).
37. D. A. Frail, F. Bertoldi, G. H. Moriarty-Schieven, et al., *ApJ* **565**, 829 (2002)
38. J. Choi, K. Nagamine, *MNRAS* **407**, Iss. 3, 1464, arXiv:0909.5425 (2010).